\documentclass{article}

\usepackage{microtype}
\usepackage{graphicx}
\usepackage{subcaption}
\usepackage{booktabs} 

\usepackage{hyperref}

\usepackage[preprint]{icml2026}

\usepackage{amsmath}
\usepackage{amssymb}
\usepackage{mathtools}
\usepackage{amsthm}

\usepackage[capitalize,noabbrev]{cleveref}

\usepackage{algorithm}
\usepackage{algorithmic}
\usepackage{xcolor}
\usepackage{graphicx}
\usepackage{subcaption}
\usepackage{caption}

\usepackage{makecell}
\usepackage{booktabs}
\usepackage[shortlabels]{enumitem}
\usepackage{wasysym}
\usepackage{float}
\usepackage{xspace}

\newcommand{\kfive}{\textsc{k5}\xspace}        
\newcommand{\kfivelite}{\textsc{k5-lite}\xspace} 
\newcommand{\modelfull}{\textsc{k6a\_5b}\xspace}
\newcommand{\modellite}{\textsc{k6av\_lite}\xspace}

\usepackage[textsize=tiny]{todonotes}

\icmltitlerunning{Adding Voice Cloning to Text-to-Audio-Video Models}

\begin{document}

\twocolumn[
  \icmltitle{Adding Voice Cloning to Text-to-Audio-Video Models with a \\ Single Zero-Initialised Layer}




  \begin{icmlauthorlist}
    \icmlauthor{Ivan Mikheev}{kandinsky}
    \icmlauthor{Viacheslav Vasilev}{kandinsky}
    \icmlauthor{Anna Dmitrienko}{kandinsky}
    \icmlauthor{Alexey Letunovskiy}{kandinsky}
    \icmlauthor{Ivan Kirillov}{kandinsky}
    \icmlauthor{Kirill Chernyshev}{kandinsky}
    \icmlauthor{Denis Dimitrov}{kandinsky}
  \end{icmlauthorlist}

  \icmlaffiliation{kandinsky}{Kandinsky Lab, Moscow, Russia}

  \icmlcorrespondingauthor{Ivan Mikheev}{ivan.mikheev@kandinskylab.ai}

  \icmlcorrespondingauthor{Viacheslav Vasilev}{viacheslav.vasilev@kandinskylab.ai}


  \vskip 0.3in
]



\printAffiliationsAndNotice{}  

\begin{abstract}
Text-to-audio-video (T2AV) generation models produce a video and its soundtrack from a textual description, but offer no control over \emph{whose voice} speaks in the output. We show that a base T2AV model can be turned into a voice-cloning model by adding a \emph{single zero-initialized linear layer} on top of its audio backbone, fine-tuning for a comparatively short training schedule, and conditioning on a short reference recording at inference time.
The reference is injected through two complementary signals: its diffusion latents are \emph{prepended} to the audio stream, and a global speaker embedding modulates token of the target audio.
On a benchmark of $674$ speaker--text pairs spanning $30$ speakers we compare against five strong voice-cloning text-to-speech baselines: our enhanced 5B model attains the highest speaker-encoder cosine similarity (SECS) across three independent verification networks (ECAPA-TDNN, WavLM-SV, Resemblyzer), statistically significantly outperforming every baseline. A side product of the architecture is that the audio path can be evaluated without the video path at inference time, yielding a $\sim$30$\times$ speed-up over the full audio-video diffusion loop while preserving the voice-cloning behaviour.
\end{abstract}

\section{Introduction}
\label{sec:intro}

Text-to-audio-video (T2AV) diffusion models~\cite{liu2024syncflow, li20253mdit, hacohen2026ltx} jointly synthesize a video clip and its soundtrack from a textual prompt. Existing systems can faithfully describe a scene, including the \emph{kind} of sound that should accompany it, but they offer no control over the \emph{identity} of the speaker that appears in the synthesized audio. A voice instruction in the prompt is realised by an arbitrary sample from the distribution trained on the data, which limits the use of T2AV models for personalized content creation, dubbing of avatar characters, or audio-visual rendering of a specific person.

A parallel line of work in text-to-speech (TTS) has produced voice-cloning systems that copy a target speaker's timbre from a short reference clip~\cite{wang2023vall, casanova2022yourtts}. These models, however, are speech-only: they cannot generate the corresponding video, and they typically rely on specialized architectures with a dedicated speaker branch. Equipping an existing T2AV foundation model with the same ability requires either a costly retraining from scratch or a careful architectural surgery that does not destroy the already-learned audio-visual prior.

In this paper we present a minimal recipe for adding \emph{reference-conditioned voice cloning} to pretrained T2AV models. Our central observation is that the two modifications -- prepending reference latents to the audio stream and modulating audio signals with a global speaker embedding through FiLM~\cite{perez2018film} -- fit naturally into the asymmetric audio–video DiT architecture. They require only a single new linear layer, which is zero-initialized so that the augmented model starts as a functional copy of the base model. We apply the approach on two our models -- \textsc{k5}\xspace (5\,B parameters) and \textsc{k5-lite}\xspace (0.6\,B parameters).


We summarize our contributions as follows:
\begin{itemize}
  \item A simple drop-in extension of a T2AV diffusion models that adds voice-cloning conditioning with one zero-initialized linear layer on top of the audio backbone.
  
  \item A short fine-tuning recipe consisting of a single voice-aware stage on top of the base T2AV backbone, which allows the model to perform voice cloning while preserving the original audio-visual generation quality.
  
  \item A three-encoder evaluation protocol for speaker similarity on a $674$-sample benchmark of unique-text speaker pairs, on which our model significantly outperforms strong text-to-speech baselines in reference speaker fidelity.
  
  \item A practical inference variant in which the audio path of the asymmetric AV-DiT is run in isolation, giving a $\sim$30$\times$ speed-up while preserving the voice-cloning behaviour of the full AV inference loop.
\end{itemize}

\section{Method}
\label{sec:method}

\subsection{Base model}
\label{sec:method:base}

\begin{figure}[ht!]
\centering
\includegraphics[width=\linewidth]{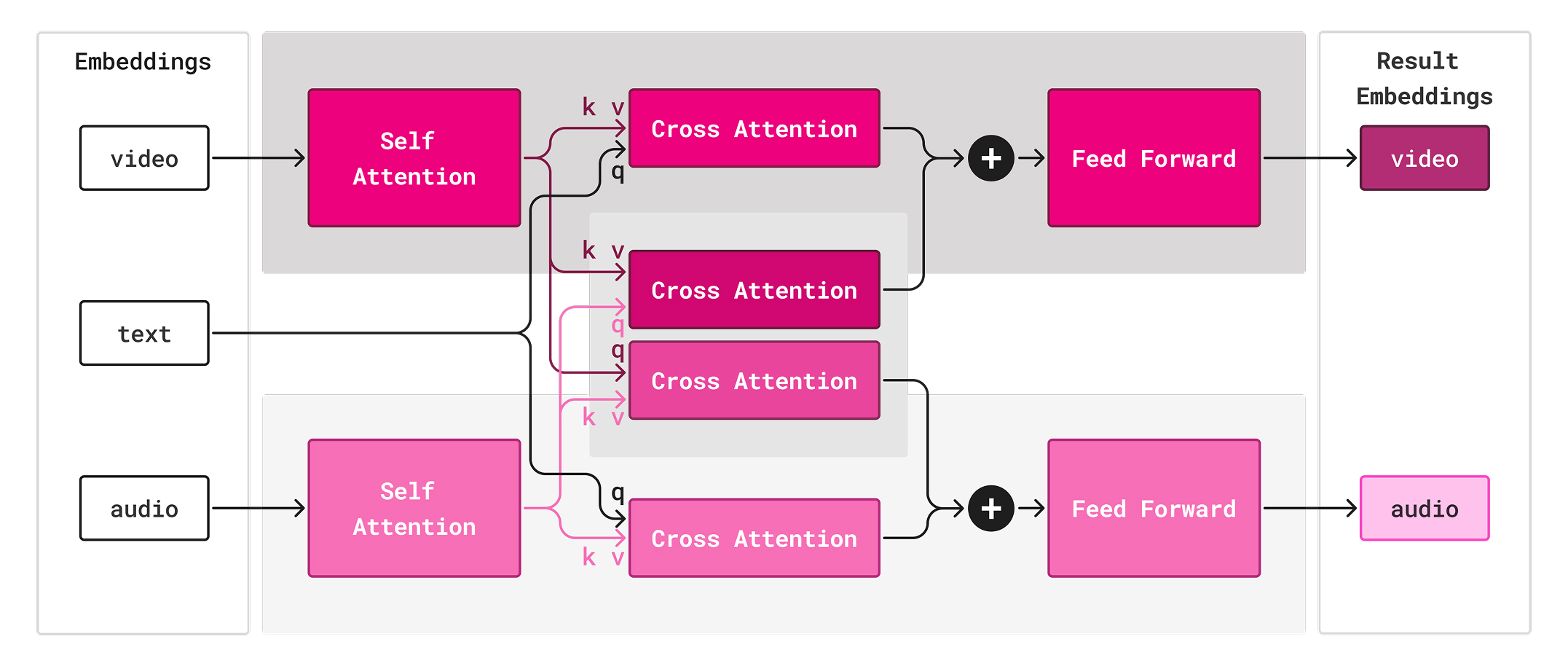}
\caption{Decoder transformer block for our AV-DiT architecture. Audio and video streams are fused using cross-attention. Each stream is based on the CrossDiT architecture~\cite{arkhipkin2026kandinsky50}.}
\label{fig:T2AVscheme}
\end{figure}

Our backbone is an asymmetric audio--video diffusion transformer (AV-DiT) with separate video and audio model dimensions ($d_v$\,=\,1792 and $d_a$\,=\,896) connected through cross-modal attention inside each fused block (Figure~\ref{fig:T2AVscheme}). We based the DiT architecture for both video and audio streams on the CrossDiT architecture from the open-source model Kandinsky 5.0~\cite{arkhipkin2026kandinsky50}. The text condition is shared between the two streams; audio latents are produced by an off-the-shelf neural audio VAE~\cite{cheng2025mmaudio} at 44.1\,kHz, and video latents by a HunyuanVideo VAE~\cite{kong2025hunyuanvideosystematicframeworklarge}. Other details about this architecture can be found in Appendix~\ref{app:architecture}. We experiment with two checkpoints of this architecture: \textbf{\kfive{}}, a 5\,B-parameter model trained on a large internal corpus of audio--video scenes, and \textbf{\kfivelite{}}, a 3\,B-parameter variant trained on the same data whose audio stream accounts for only $\approx 0.6$\,B parameters. For brevity we call the augmented models, after fine-tuning, \textbf{\modelfull} (the 5\,B checkpoint, full audio--video inference) and \textbf{\modellite} (the 3\,B checkpoint run in the audio-only inference mode of Section~\ref{sec:audioonly}, which evaluates only its $\approx 0.6$\,B audio sub-network).


\subsection{Reference-conditioned voice cloning}
\label{sec:method:ref}

We expose two complementary reference signals to the audio backbone of the AV-DiT.

\paragraph{Prepended reference latents.} Given a target audio sequence $x_a = \mathrm{VAE}(a_\mathrm{target}) \in \mathbb{R}^{T_a\times d_a}$ and a reference recording $a_\mathrm{ref}$ of $1.2\text{--}4$ seconds taken from the same scene during training, we encode the reference with the \emph{same} audio VAE and \emph{prepend} the resulting latents to the audio stream:
\begin{equation}
\tilde{x}_a \;=\; [\,\mathrm{VAE}(a_\mathrm{ref}) \,\Vert\, x_a\,]
\;\in\; \mathbb{R}^{(T_\mathrm{ref}+T_a)\times d_z}.
\label{eq:prepend}
\end{equation}
Throughout the 32 fused decoder blocks, the target tokens attend to the reference tokens via the same self-attention mechanism that already exists in the base AV-DiT (Figure~\ref{fig:T2AVscheme}), so the model acquires reference conditioning without any new attention layers. At inference, the diffusion update is restricted to the target portion of the composite sequence; the reference portion is held fixed so that the model is forced to keep its timbre consistent across all denoising steps.



\paragraph{Speaker FiLM.}
Long-range timbre information that is not easily captured by local self-attention is provided through a global speaker embedding $e \in \mathbb{R}^{1024}$ obtained from a frozen Qwen3-TTS speaker encoder~\cite{hu2026qwen3ttstechnicalreport}. A single new linear layer $W_\mathrm{film}: \mathbb{R}^{1024}\!\to\!\mathbb{R}^{2d_a}$ maps this embedding to scale and shift coefficients $(\gamma, \beta)$ that modulate each audio token of the \emph{target} portion via FiLM~\cite{perez2018film}:
\begin{equation}
h_a \;\leftarrow\; h_a \odot (1+\gamma) + \beta,
\qquad (\gamma,\beta) = W_\mathrm{film}(e).
\label{eq:film}
\end{equation}
Reference tokens are not modulated. The weights and bias of $W_\mathrm{film}$ are initialized to zero, so $(\gamma,\beta)=(0,0)$ at the start of fine-tuning and the augmented model produces \emph{exactly} the same outputs as the base T2AV model on its first forward pass.


\subsection{Classifier-free guidance with two directions}
\label{sec:method:cfg}

During training, with independent probabilities of $0.1$ each we drop the text condition and the reference signal (latents and speaker embedding jointly), giving the model the four configurations required for a three-way split classifier-free guidance at inference~\cite{ho2021cfg}:
\begin{equation}
\hat{\epsilon} \;=\; \epsilon_\varnothing
+ w_t\,(\epsilon_\mathrm{text} - \epsilon_\varnothing)
+ w_r\,(\epsilon_\mathrm{full} - \epsilon_\mathrm{text}),
\label{eq:splitcfg}
\end{equation}
where $\epsilon_\varnothing$ refers to unconditional generation, $\epsilon_\mathrm{text}$ -- text-conditioned generation and $\epsilon_\mathrm{full}$ -- generation with condition to text and reference audio. The second term controls the prompt adherence strength, and the third controls the reference voice imposition strength. This split lets us trade text fidelity and speaker fidelity at inference time without retraining.


\subsection{Training schedule}
\label{sec:method:train}

We fine-tune the entire model -- $W_{\mathrm{film}}$ together with the existing audio backbone weights -- on top of the base T2AV checkpoint using a single \emph{voice-aware} stage. We use AdamW optimizer with $1\!\times\!10^{-5}$ for the base parameters and $5\!\times\!10^{-5}$ for $W_{\mathrm{film}}$, and sample the reference window from speech-rich segments of each clip after the target portion. In 10\% of training steps, we independently feed unnoised audio and video latents to retain the original AV joint distribution. That is, the model remains a T2AV generator first and a voice-cloner second. Implementation details can be found in Appendix~\ref{sec:suppl:impl}.


\section{Experiments}
\label{sec:experiments}





\subsection{Benchmark and metrics}
\label{sec:exp:bench}

We evaluate on a voice-cloning benchmark built from the VCTK corpus~\cite{yamagishi2019vctk} at its native $48$\,kHz studio quality: $674$ samples spanning $30$ native-English speakers, where each speaker is assigned its \emph{own} sentences ($100\%$ unique texts, $5$--$16$ words) rather than a shared passage, so that intelligibility and speaker similarity are not inflated by text overlap. Each sample provides one reference clip, three additional enrollment clips of the same speaker, and a held-out target text.

For a fair comparison across architectures, we feed each reference to every model in its own native sample rate, apply an identical pre-processing pipeline before all metrics (silence trimming, peak-normalization to $-3$\,dBFS, downmixing to $16$\,kHz mono), and evaluate \emph{all} systems at the same $16$\,kHz rate so that audio bandwidth cannot bias any metric. As $\modelfull$ and $\modellite$ are diffusion models with no explicit target-length conditioning, we set the generated duration adaptively from the prompt word count ($\mathrm{seconds}=\mathrm{words}/2.6+1.3$, clamped to $[4,12]$\,s); the text-to-speech baselines determine their own output length.

\paragraph{Baselines.} We compare against five external voice-cloning text-to-speech systems: a $1.7$\,B-parameter Qwen3-TTS~1.7B and its $0.6$\,B counterpart Qwen3-TTS~0.6B~\cite{hu2026qwen3ttstechnicalreport}, the autoregressive XTTS-v2~\cite{casanova2024xtts}, the auto-regressive IndexTTS2~\cite{zhou2025indextts2}, and the public NAVA checkpoint~\cite{ji2026nava}, a $6.3$\,B joint audio--video diffusion model based on an Align-then-Fuse MMDiT architecture with reference timbre control provided by Timbre-in-Context Conditioning.

\paragraph{Metrics.}
We report speaker-encoder cosine similarity (SECS) using three complementary verification networks: \textbf{ECAPA-TDNN}~\cite{desplanques2020ecapa}, \textbf{WavLM-base-plus-sv}~\cite{chen2022wavlm}, and \textbf{Resemblyzer / GE2E}~\cite{wan2018ge2e}.
For each network we compute two similarities that differ only in what the generated utterance is compared against. The \emph{vs.\ reference} score is the cosine similarity between the generated embedding and the \emph{single} reference clip given to the model. The \emph{vs.\ centroid} score instead compares the generated embedding against the \emph{centroid} of four clips of the same speaker (the reference plus three enrollment clips), which reduces the variance of a single-utterance reference and gives a more stable estimate of the target speaker identity. Intelligibility is measured by word and character error rates \textbf{WER\%}/\textbf{CER\%} of Whisper transcriptions \cite{radford2023whisper}, and by \textbf{WER$_0$\%}, the fraction of samples transcribed with zero word errors.

\subsection{Main results}
\label{sec:exp:results}

Table~\ref{tab:ruen} reports the comparison on the full
$674$-sample benchmark.
\textbf{$\modelfull$} attains the highest speaker similarity on
\emph{every} one of the six SECS columns, outperforming all six
competing systems on both the reference-based and the
centroid-based scores across all three verification networks.
The lead holds on the two most reliable encoders, WavLM-SV
($0.944$ vs.\ reference) and Resemblyzer ($0.866$), as well as on
ECAPA-TDNN ($0.766$); averaged over the three networks the margin
over the strongest external baseline (Qwen3-TTS~0.6B) is $0.041$
and is confirmed by a paired Wilcoxon signed-rank test
($p<10^{-89}$ against each competitor).

The Qwen3-TTS and IndexTTS2 baselines achieve very low WER/CER, but
do so by re-rendering each prompt with a clean, studio-like voice
that is more distant from the actual reference speaker (lower SECS).
$\modelfull$ and $\modellite$ make a different trade-off:
they faithfully imitate the reference, including its acoustic
quirks, at the cost of a higher WER inherited from the underlying
T2AV prior. We attribute the stronger speaker fidelity to the
generative audio prior of the base T2AV model: rather than mapping
text to a normalized voice as autoregressive TTS systems tend to do,
the diffusion backbone reconstructs fine-grained acoustic detail of
the prepended reference, at the price of occasional hallucinated
words that inflate WER on short or difficult prompts.


\begin{table*}[t]
\centering
\caption{English VCTK benchmark ($674$ samples per model,
$30$ speakers, unique texts).
The SECS columns are speaker-encoder cosine similarities
(higher is better) for three verification networks
(E: ECAPA-TDNN, W: WavLM-SV, R: Resemblyzer), reported both
against the single reference clip (\emph{vs.\ reference}) and
against the centroid of the reference and three enrollment clips
(\emph{vs.\ centroid}).
WER\%/CER\% are word/character error rates (lower is better) and
WER$_0$\% is the fraction of samples transcribed with zero word
errors (higher is better). Best value per column in \textbf{bold},
second best \underline{underlined}.}
\label{tab:ruen}
\setlength{\tabcolsep}{1.5pt}
\begin{tabular}{lrrrrrrrrr}
\toprule
 & \multicolumn{3}{c}{SECS vs.\ reference} & \multicolumn{3}{c}{SECS vs.\ centroid} & & & \\
\cmidrule(lr){2-4}\cmidrule(lr){5-7}
Model & E & W & R & E & W & R & WER\% & CER\% & WER$_0$\% \\
\midrule
$\modelfull$ (ours)     & \textbf{0.766} & \textbf{0.944} & \textbf{0.866}
                 & \textbf{0.770} & \textbf{0.951} & \textbf{0.878}
                 & 5.76 & 5.21 & 86.6 \\
Qwen3-TTS 0.6B & 0.678 & 0.936 & 0.840
                 & \underline{0.711} & 0.946 & 0.864
                 & \phantom{0}\underline{0.81} & \phantom{0}0.36 & \textbf{95.7} \\
Qwen3-TTS 1.7B & 0.674 & \underline{0.938} & 0.839
                 & 0.709 & \underline{0.951} & 0.864
                 & \phantom{0}\textbf{0.74} & \phantom{0}\textbf{0.17} & \underline{95.3} \\
IndexTTS2        & \underline{0.695} & 0.885 & \underline{0.854}
                 & 0.693 & 0.878 & \underline{0.867}
                 & \phantom{0}0.96 & \phantom{0}\underline{0.22} & 93.2 \\
XTTS-v2         & 0.575 & 0.923 & 0.816
                 & 0.611 & 0.940 & 0.841
                 & \phantom{0}1.14 & \phantom{0}0.32 & 93.2 \\
$\modellite$ (ours)   & 0.640 & 0.847 & 0.826
                 & 0.630 & 0.846 & 0.843
                 & 5.08 & 3.38 & 69.4 \\
NAVA             & 0.624 & 0.852 & 0.759
                 & 0.625 & 0.850 & 0.780
                 & 5.82 & 3.21 & 69.1 \\
\bottomrule
\end{tabular}
\end{table*}

\subsection{No regression of the base audio model}
\label{sec:audioonly:noregress}


We verify that adding voice-cloning conditioning does not degrade the
reference-free generation ability of the pretrained audio backbone.
We compare the base T2AV checkpoint against \emph{itself after our
reference-aware fine-tuning}, running \emph{both} models in the
reference-free regime (text$\rightarrow$audio only, no reference clip)
on $400$ prompts covering $30$ speakers, with identical seeds and
texts. A regression would show up as the fine-tuned model moving
\emph{away} from real speech or losing intelligibility. We measure
Fr\'echet Audio Distance (FAD) against real speech, word/character
error rates (WER/CER), the fraction of perfectly transcribed samples
(WER$_0$), CLAP text--audio alignment, and UTMOS naturalness.

Table~\ref{tab:noregress} reports the \emph{relative} change of the
fine-tuned model with respect to the base model. Rather than
regressing, the fine-tuned checkpoint improves on every objective
axis: it is substantially closer to the distribution of real speech
(FAD), roughly halves the transcription error and more than doubles
the fraction of perfectly transcribed samples, aligns better with the
prompt (CLAP), and matches perceptual naturalness (UTMOS). We
attribute this to the zero-init design of $W_\mathrm{film}$
(Sec.~\ref{sec:method:ref}): when the reference is absent the FiLM
path stays close to identity, so the augmented model remains a strict
functional superset of the base T2AV generator while benefiting from
extra in-domain updates.

\begin{table}[t]
\centering
\caption{No-regression check: \emph{relative} change of the base
T2AV model \emph{after} reference-aware fine-tuning, evaluated in
the reference-free regime on $400$ prompts ($30$ speakers, identical
seeds/texts). A positive change is an improvement for
$\uparrow$-metrics and a negative change is an improvement for
$\downarrow$-metrics; every metric moves in the improving direction.}
\label{tab:noregress}
\small
\setlength{\tabcolsep}{6pt}
\begin{tabular}{lr}
\toprule
Metric & Relative change \\
\midrule
FAD vs.\ real speech $\downarrow$   & $-30.6\%$ \\
WER $\downarrow$                    & $-46.0\%$ \\
CER $\downarrow$                    & $-47.0\%$ \\
WER$_0$ (perfect samples) $\uparrow$ & $+113\%$ \\
CLAP text--audio $\uparrow$         & $+29.4\%$ \\
UTMOS naturalness $\uparrow$        & $+0.6\%$ \\
\bottomrule
\end{tabular}
\end{table}

\subsection{Human side-by-side study}
\label{sec:suppl:sbs}

In addition to the objective no-regression check
(Sec.~\ref{sec:audioonly:noregress}, Table~\ref{tab:noregress}),
we run a human side-by-side (SBS) study to confirm that
reference-aware fine-tuning does not degrade the perceived quality
of reference-free generation. We compare the base audio model
against \emph{itself after our reference-aware fine-tuning} on
$100$ held-out text-to-audio prompts, with both models generating
\emph{without} a reference recording, so that any regression would
show up as a lower win rate for the fine-tuned model. Annotators
rate each pair on prompt following, technical quality, and
speech/aesthetic quality as either a win for one of the two models
or a draw. Table~\ref{tab:noregress_sbs} shows that the fine-tuned
checkpoint scores marginally higher win rates ($52$--$57\%$) on all
three axes, consistent with the objective results. A smaller SBS
study on the full T2AV pipeline (reference-free) shows the same
pattern, so the ``strict superset'' property carries over from
audio-only to full multimodal.

\begin{table}[t]
\centering
\caption{SBS human evaluation on $100$ held-out text-to-audio
prompts, base audio model vs.\ the same model after
reference-aware fine-tuning (both generate \emph{without} a
reference). Numbers are percentages of pair-wise judgements
per axis.}
\label{tab:noregress_sbs}
\small
\setlength{\tabcolsep}{4pt}
\begin{tabular}{lccc}
\toprule
                            & Prompt      & Technical & Speech/Aesth.\\
                            & follow.\    & quality   & quality      \\
\midrule
Base wins                   & \phantom{0}9  & \phantom{0}8  & 11 \\
Fine-tuned wins             & 14 & 23 & 17 \\
Both good                   & 78 & 75 & 78 \\
Both bad                    & \phantom{0}9  & \phantom{0}4  & \phantom{0}4  \\
\midrule
Base win rate               & 48 & 43 & 47 \\
Fine-tuned win rate         & \textbf{52} & \textbf{57} & \textbf{53} \\
\bottomrule
\end{tabular}
\end{table}
\subsection{Fast Audio-only Inference from an AV Checkpoint}
\label{sec:audioonly}

The asymmetry of the AV-DiT lets us run the audio path \emph{in
isolation} at inference without changing any weights, by
short-circuiting the video sub-block of each fused decoder block and
skipping the video VAE decoder (details in
Appendix~\ref{sec:suppl:audioonly}). This drops the cost of one
diffusion step from a $3.18$\,B-parameter joint pass to an effective
$\sim$$0.58$\,B audio-only pass, a $\sim$$30\times$ speed-up.
The speed-up is not bought at the cost of voice cloning: in
Table~\ref{tab:ruen}, $\modellite$ preserves the reference timbre
well, at a modest SECS drop of $\approx0.09$ vs.\ the full
$\modelfull$, which we read as the video signal acting as a mild
regularizer on the audio path. In practice, $\modellite$ enables
quick auditioning of a reference across many prompts before
committing to full $\modelfull$ rendering.

\section{Conclusion}
\label{sec:conclusion}

We have shown that T2AV diffusion models can be turned into voice-cloning systems with a remarkably small architectural change: prepending the reference latents to the audio stream and adding a single zero-initialized linear layer that drives speaker FiLM on the audio backbone, trained in a short voice-aware stage. On a $674$-sample benchmark our $\modelfull$ obtains the highest speaker similarity across three verification networks, significantly outperforming every baseline, while the audio-only variant $\modellite$ preserves this behavior at a $\sim$30$\times$ lower inference cost and the original text-to-audio-video functionality stays intact.

Two extensions are natural: the prepend+FiLM recipe should transfer to any asymmetric AV-DiT paired with a frozen speaker encoder, and the WER gap to dedicated TTS suggests adding a small text-fidelity loss (e.g.\ CTC against the prompt) without disturbing the speaker conditioning.


\bibliography{references}
\bibliographystyle{icml2026}

\newpage
\appendix
\onecolumn
\clearpage
\section{Implementation details}
\label{sec:suppl:impl}

\subsection{Base architectures}\label{app:architecture}
\kfive{} and \kfivelite{} share the same asymmetric AV-DiT
backbone described in Section~\ref{sec:method:base} (Figure \ref{fig:T2AVscheme}). Both consist of 32 fused decoder blocks, with a video stream of width $d_v=1792$ (heads of dimension 64) and an audio stream of width $d_a=896$ (heads of dimension 64). A 2-block text encoder consumes the prompt and feeds it to both
streams. $\kfive$ uses the full 32 fused blocks, $\kfivelite$ a reduced configuration with the same block topology but smaller text and output layers, totaling $\approx 3$\,B parameters, of which the audio stream (the part evaluated in the audio-only inference mode) accounts for only $\approx 0.6$\,B.

\subsection{Reference window sampling}
At training time, given a training clip, we sample the target
audio from the first slice of the parquet record (typically 5
seconds of audio aligned with the video latents).
The reference window is drawn from the \emph{remaining} audio
of the same clip, with a buffer of $0.5$\,s between target and
reference to avoid trivial copy.
Speech references are sampled with duration in $[1.2,\,2.5]$\,s
from a 12\,s search window, non-speech references with duration
in $[1.5,\,3.5]$\,s from a 3.5\,s search window.
A minimum-RMS and a maximum-clipping-ratio quality filter discards
silent or saturated references.

\subsection{Hyperparameters}
\begin{itemize}\itemsep0pt
\item Optimizer: AdamW, $(\beta_1,\beta_2)=(0.9,0.95)$,
      weight decay $10^{-3}$, max grad norm $1.0$.
\item Base LR: $10^{-5}$; LR multiplier for the new linear layer
      $W_\mathrm{film}$ and its biases: $5\times$.
\item Warmup: $8000$ steps, constant afterwards.
\item Reference-conditioning drop probability: $0.1$;
      text-conditioning drop probability: $0.1$.
\item Modality-clean schedule: $10\%$ of steps run with the video
      latent clamped to zero (audio-only loss) and $10\%$ with
      the audio latent clamped to zero (video-only loss), to
      retain joint AV behavior throughout fine-tuning.
\item Default inference: $50$ diffusion steps, split CFG
      with $w_t=5$ and $w_r=4$, fixed seed.
\end{itemize}

\subsection{Speaker encoder}
The frozen speaker encoder used to compute the FiLM~\cite{perez2018film} embedding is a publicly released Qwen3-TTS speaker network~\cite{hu2026qwen3ttstechnicalreport} operating on $24$\,kHz audio. We run it once per training sample on CPU in a background worker and only transmit the resulting 1024-dim vector to the training step.

\section{Audio-only inference details}
\label{sec:suppl:audioonly}

Because the AV-DiT is asymmetric, the same fine-tuned checkpoint can
be evaluated with the video stream disabled, without any weight
changes (Sec.~\ref{sec:audioonly}). Concretely, inside each of the
$32$ fused decoder blocks the video sub-block is short-circuited: its
$1792$-dim video self-attention and FiLM modulation are never
computed, while the audio sub-block, the speaker FiLM layer, and the
audio classifier-free-guidance loop keep operating exactly as in
training. The video VAE decoder is likewise skipped. This reduces the
per-step forward from a $3.18$\,B-parameter joint-modality pass to an
effective $\sim$$0.58$\,B audio-only pass, giving the
$\sim$$30\times$ speed-up reported in the main text.
The audio-only variant $\modellite$ trades a small amount of speaker
similarity ($\approx0.09$ in SECS averaged over the three encoders)
for this speed-up; we attribute the gap to the loss of the video
stream, which otherwise acts as a mild regularizer on the audio path.

\section{Additional analysis}
\label{sec:suppl:ablation}

We carried out a more extensive factor study to identify which
inference-time and data-time choices matter most for voice
cloning quality.
The following findings are based on a 82-run sweep across a
held-out subset of the benchmark.

\paragraph{Reference length.}
Speaker similarity grows monotonically with reference length
from $\sim 1$\,s up to a plateau at $4$\,s, beyond which
returns are negligible.
Below $2$\,s the Resemblyzer SECS drops sharply,
suggesting that the speaker encoder rather than the diffusion
model is the bottleneck on very short references.

\paragraph{Reference-guidance weight.}
The optimum of the second CFG weight $w_r$
(Eq.~\ref{eq:splitcfg}) is in the range $4$--$6$.
Below $3$ the reference is under-imposed; above $7$ we
observe a small over-smoothing of the audio that reduces UTMOS.

\paragraph{Number of diffusion steps.}
The mean SECS is essentially saturated by $30$ steps;
moving from $30$ to $60$ adds $<0.02$ in mean SECS
and is not worth the doubled wall-clock cost for routine use.

\paragraph{Language matching.}
Generating from a prompt whose language differs from the
reference language is possible but lowers the mean SECS
by about $0.05$ on average.
We therefore recommend matching prompt and reference languages
whenever possible.

\paragraph{Reference pre-processing.}
A simple de-noise / VAD pre-processing of the reference
recording is helpful on noisy or amateur captures but is
harmful on already-clean studio references, where it can
slightly hurt SECS.
A safe default is to skip pre-processing when the reference
already has UTMOS $\gtrsim 3$.


\end{document}